# Heavy Electron doping-induced antiferromagnetic phase as the parent for iron-oxypnictide superconductor LaFeAsO$_{1-x}$H$_x$

Soshi Iimura[1], Satoru Matsuishi[2], and Hideo Hosono[1, 2]

[1] Laboratory for Materials and Structures, Tokyo Institute of Technology, Yokohama 226-8503, Japan

[2] Materials Research Center for Element Strategy, Tokyo Institute of Technology, Yokohama 226-8503, Japan

Correspondence should be addressed to H. Hosono (email: hosono@msl.titech.ac.jp)

# I. ABSTRACT

We perform transport measurements and band structure calculations of electron-doped LaFeAsO$_{1-x}$H$_x$ over a wide range of $x$ from 0.01 to 0.66. The $T^2$ and $\sqrt{T}$ dependency of the resistivity are observed at $x \sim 0.17$ and 0.41, respectively. The sign change of $R_H$ without opening of the spin-density-wave gap for $0.45 \leq x \leq 0.58$ and $T < T_N$ as well as the calculated non-nested Fermi surface at $x = 0.5$ indicate the more localized nature of the AF2 as compared to spin density wave phase at non-doped sample. Considering the results from band calculations and the finite size of Hund's rule coupling, the change of the normal conducting state with $x$ is reasonably explained by a strong depression of the coherent scale owing to the increased effective Coulomb repulsion in the narrow anti-bonding $3d_{xy}$ band that approaches the half-filled regime. The following transition from the paramagnet to AF2 is understood as the quenching or ordering of the less-screened spins with a large entropy at low temperature. These results suggest that the normal conducting properties over a wide doping range for the LaFeAsO$_{1-x}$H$_x$ are strongly influenced by the local spins in the incoherent region. Thus, we conclude that the parent phase in LaFeAsO$_{1-x}$H$_x$ is not the spin-density wave but the AF2, which may be primarily responsible for the singular superconducting properties of the 1111 type compared with other iron pnictides.

## II. INTRODUCTION

Iron pnictides and chalcogenides are the second class of high-critical temperature ($T_c$) superconductors and are characterized by their wide variety of crystal structures [1,2]. A $T_c$ exceeding 50 K is achieved for the *Ln*FeAsO series (*Ln* = lanthanides), termed as 1111 type, by doping electrons into the parent phase with anti-ferromagnetism (AFM) [3–5].

Despite the similarity of the phase diagrams of cuprates and iron pnictides, the electronic states of both parent compounds are strikingly different; the former is an AFM Mott insulator with strong electron correlations from the on-site Coulomb repulsion, $U$, in the single $3d$ band of the cupric ion, whereas the latter shows metallic conduction with a spin-density-wave (SDW) transition at low temperatures [6]. From the discovery of the iron pnictide superconductors, the non-negligible electron correlations have been debated because the resistivity and spin susceptibility in the normal states are strongly enhanced from the values of conventional metals [7,8]. In a pioneering article by Haule and Kotliar that reported the properties of the electron-doped *Ln*FeAsO$_{1-x}$F$_x$, a Hund's rule coupling was proposed to indeed be responsible for the correlation [8]. They introduced a characteristic temperature, coherence scale $T^*$, and interpreted the anomalous properties as a crossover phenomenon at the $T^*$ from the Fermi liquid states with well-screened local magnetic moments by the itinerant electrons ($T < T^*$) into an incoherent bad metal with bare local moments ($T > T^*$). Currently, the theoretical framework is applied to the other compounds including iron chalcogenides, and experimentally proved by Hardy *et al* in 122-type KFe$_2$As$_2$ [9]. As for the size of the moment in the AFM phase of non-doped sample, the importance of the occupation number of the Fe-$3d_{xy}$ orbital (we choose the $x$ and $y$ axes in the Fe-Fe direction, hereafter) is indicated, and a clear inversely proportional relationship between the occupation and the moment size is found across all iron pnictides and chalcogenides [10]. According to this result, the large moment of FeTe is rationally explained by the increased effective Coulomb repulsion $U_{eff}$, which is a consequence of the reduced occupation number approaching a half-filled regime [11].

Recently, we observed another AFM phase (AF2) accompanying a unique structural transition from tetragonal to orthorhombic in the range of $0.40 \leq x \leq 0.51$ in LaFeAsO$_{1-x}$H$_x$ [12]. It should be noted that the maximum magnetic moment, 1.21 μ$_B$ at $x$ = 0.51, is twice as large as that of AFM (AF1) in the non-doped sample (0.63 μ$_B$), the largest in iron pnictides. Additionally, in the proximity of the AF2, the higher-$T_c$ superconductivity (SC2) at 36 K is observed for $0.20 \leq x \leq 0.45$ compared with the maximum $T_c$ = 29 K found previously in the conventional superconducting phase (SC1) for $0.05 \leq x \leq 0.20$ [13]. This pronounced magnetic moment in AF2 cannot be understood using the above theory because the occupation number of the 3$d$ orbital is significantly increased by heavy electron doping through substitution of H$^-$ into the O$^{2-}$ site (O$^{2-} \rightarrow$ H$^-$ + $e^-$). Nevertheless, the enhanced $T_c$ in SC2 appears to be reminiscent of the non-negligible relation between the large magnetic moment (or the strong electronic correlation) and the pairing mechanism.

Herein, we present a systematic investigation of electron-doped LaFeAsO$_{1-x}$H$_x$ with a large $x$ ranging from 0.01 to 0.66 by means of transport measurements and band structure calculations. In Sec. IV-A, we provide experimental results of the resistivity and Hall effect measurements of LaFeAsO$_{1-x}$H$_x$. From the resistivity analyses, the crossover temperature, $T_{CO}$, is detected as the inflection points of the temperature dependency of resistivity in the normal states for $0.05 \leq x < 0.41$. After suppression of the AF1, the $T_{CO}$ gradually increases as $x$ increases, and reaches a maximum value of 214 K at $x$ = 0.17, below which the Fermi liquid state is developed. Further doping reduces the $T_{CO}$ dramatically, and finally becomes unobservable at $x$ = 0.41, where a $\sqrt{T}$ dependency of resistivity, often referred to as a characteristic of the incoherent region, is observed. Just above the $T_{CO}$, the temperature dependence of the Hall coefficient, $R_H$, is also affected by the strong electron-spin scattering in the incoherent region. Finally, for $x \geq 0.45$, the $R_H$ shows the sign change at the transition temperature into the AF2 without the opening of the SDW gap, indicating that the AF2 is not described by the itinerant picture but rather by the more localized magnetism. In Sec. IV-B and C, we examine the band structures of LaFeAsO$_{1-x}$H$_x$ to understand the experimental results. By introducing molecular orbitals composed of two atoms inside the unit cell and taking into account the symmetry requirements, we found that an in-phase Fe-3$d_{xy}$ orbital forms a single narrow band lying on the Fermi level ($E_F$), whereas the out-of-phase Fe 3$d_{xy}$ orbitals provide two dispersed bands by hybridizing with the As-4$p_Z$ orbital. Although these bands are highly entangled at $x = 0$, the electron doping leads to separation of the bands into the destabilized in-phase Fe-3$d_{xy}$ orbital and the stabilized out-of-phase orbital located below $E_F$. Consequently, the narrow in-phase Fe-3$d_{xy}$ orbital becomes almost half-filled at $x = 0.75$. These calculated results tell us that the magnetic moments in the half-filled Fe-3$d_{xy}$ orbital are responsible for the incoherent transport properties and the following magnetic ordering because the screening of the magnetic moments in a half-filled orbital is highly restricted by the Hund's rule coupling. It should be noted that such an electron-doping-induced electron correlation characterizing the 1111 type has not been observed in the other iron pnictides, for example Ba(Fe$_{1-x}$Co$_x$)$_2$As$_2$; the non-Fermi liquid state as well as superconductivity emerge in proximity of the SDW phase, and are monotonically diminished for $x > 0.15$, where the Fermi liquid state is developed instead [14]. These experimental and calculated results suggest that in the LaFeAsO$_{1-x}$H$_x$, the normal and superconducting states are strongly influenced by the local moment in the wide doping region except for the non-doped case. Therefore, we conclude that the parent phase in the LaFeAsO$_{1-x}$H$_x$ is not the SDW but the AF2, which may primarily distinguish the physical properties of the 1111 type from other iron pnictides.

## III. EXPERIMENTAL

Polycrystalline LaFeAsO$_{1-x}$H$_x$ samples were synthesized by solid-state reaction at 1473 K under pressures of 2 and 5 GPa for samples in the range of $0.01 \leq x \leq 0.35$ and $0.41 \leq x \leq 0.66$, respectively. Chemical compositions, except for hydrogen, were measured by an electron probe micro-analyzer (JEOL JXA-8530F). The hydrogen content was determined by the thermal gas desorption spectrometry (ESCO TDS-1400TV). Transport measurements were performed using a physical properties measurement system (Quantum Design, Inc. PPMS). The electrical resistivity, $\rho_{xx}$, was measured using a four-probe technique. The Hall resistivity, $\rho_{xy}$, was measured by a DC technique under a magnetic field swept from $-9$ to $9$ T. The $R_H$ values were obtained by fitting the field dependence of $\rho_{xy}(H)$ to an equation, $\rho_{xy}(H) = R_H H + \beta H^3 + \gamma$ [15]. Except for the temperature region below the magnetic transition at $x = 0.01, 0.03, 0.52,$ and $0.58$, the non-linear component $\beta$ was negligible (See the Supplemental Material SM1 [16]).

Usually, the transport properties of polycrystalline sample have some uncertainty due to the presence of, *e. g*., the grain-boundary, impurity phase, and anisotropy. However, utilizing the high pressure synthesis method, the sintering density of LaFeAsO$_{1-x}$H$_x$ samples result in nearly 100 %, and even at $x = 0.66$ which is the end-member of our samples, the weight percent of the 1111 phase exceeds 98 % estimated by X-ray diffraction (See the Supplemental Material SM2 [16]). In addition, because the resistivity-anisotropy of 1111-type is relatively higher than that of other types [17], the averaged resistivity of polycrystalline sample primarily reflects the resistivity in *ab*-plane.

First-principles band structures were obtained with the WIEN2k code [18], where experimentally determined structural parameters at room temperature were adopted. For the calculation of LaFeAsO$_{0.25}$H$_{0.75}$, the $\sqrt{2} \times \sqrt{2} \times 1$ supercell was used. For the generalized gradient approximation (GGA), PBEsol formulated by the Perdew, Burke, and Ernzerhof was adopted [19]. To ensure convergence, the linearized augmented plane wave basis set was defined by the cutoff $R_{MT}K_{MAX} = 9.0$ ($R_{MT}$: the smallest atomic sphere radius in the unit cell) for calculations of LaFeAsO, and $R_{MT}K_{MAX} = 4.5$ for calculations of LaFeAsO$_{0.5}$H$_{0.5}$ and LaFeAsO$_{0.25}$H$_{0.75}$. The mesh samplings in the Brillouin zone were of $10\times10\times6$ for $x = 0$ and $0.5$, and of $8\times8\times6$ for $x = 0.75$. Maximally localized Wannier functions (MLWF) as projection functions were constructed from Fe $3d$, As $4p$, O $2p$, and H $1s$ states within an energy window from $\sim -6$ to $\sim 2$ eV [20] (See the Supplemental Material SM3 [16]). As for the estimation of the hopping parameters, we used only the ten $3d$ orbitals of Fe for the projection.

## IV. RESULTS & DISCUSSION

### A. Experimental transport properties of LaFeAsO$_{1-x}$H$_x$

We experimentally investigated the resistivity and Hall effect of LaFeAsO$_{1-x}$H$_x$ with $x$ = 0.01–0.66. The temperature dependence of resistivity [$\rho_{xx}$ ($T$)] in the $x$-ranges of $0.01 \leq x \leq 0.17$ and $0.30 \leq x \leq 0.66$ are separately shown in Fig. 1(a) and (b), respectively. The resistivity in all samples decreases as the temperature is lowered, indicating a metallic nature. In Fig. 1(a), a kink observed at $x = 0.01$ and $T \sim 150$ K arises from structural and/or magnetic transitions into orthorhombic AF1 phases [21]. After the anomaly is suppressed, the superconducting transition appears for $x \geq 0.05$. The $T_c$ has a maximum value of 36 K at $x = 0.35$ and 0.37, and then is reduced for $x \geq 0.41$ [Fig. 1(b)]. In the over-doped region, the kink appears again at $T \sim 100$ K in the four samples with $x = 0.45$–0.66. Figure 1(c-e) shows the temperature differentials of the $\rho_{xx}$ ($T$). From previous research on the non-doped compound, it is known that the temperature at which the derivative starts to increase agrees with the structural transition temperature ($T_s$), while the peak-top temperature corresponds to the Neel temperature, $T_N$ [21]. As shown in Fig. 1(c), according to this empirical rule, we can estimate the $T_s$ and $T_N$ as 159 (151) and 138 (105) K for $x = 0.01$ (0.03) indicated by a filled triangle ($T_{anom.1}$) and an inverse triangle ($T_{anom.2}$), respectively. In the same manner, we determine the temperatures of the highly doped samples with $x = 0.52$, 0.58, and 0.66 as $T_{anom.1} = 101$, 114, and 116 K, and $T_{anom.2} = 93$, 100, and 106 K, respectively. The results are shown in Fig. 1(d). The $T_{anom.1}$ and $T_{anom.2}$ at $x = 0.52$ coincide well with the $T_s = 95$ and $T_N = 92$ K at $x = 0.51$, respectively [12], suggesting that these estimations would also be applicable to the ordered phase in the heavily doped region. In Fig. 1(e), the derivatives of the superconducting samples with $x = 0.05$–$0.41$ are shown. Although the structural and magnetic transitions are completely suppressed in this $x$ range, an inflection point ($T_{CO}$) denoted by the filled diamond is observable for $0.05 \leq x \leq 0.37$. As $x$ increases from 0.05 to 0.17, the $T_{CO}$ increases, but decreases for $0.17 < x \leq 0.37$, and finally disappears at $x = 0.41$.

Figure 2(a–c) depicts the magnetic field dependence of $\rho_{xy}$ $(H)$ at $T = 50$ K. The $\rho_{xy}$ $(H)$ shows a non-linear behavior below the $T_{NS}$ of AF1 and AF2, while a linear behavior for $0.05 \leq x \leq 0.41$ over the whole temperature range. At $x = 0.45$ and 0.52, the slopes of the $\rho_{xy}$ $(H)$ becomes positive, whereas at $x = 0.58$ and 0.66, the sign is shown to be negative. To see the temperature variation of the slope of the $\rho_{xy}$ $(H)$, we plot the linear response, $R_H$, against temperature in Fig. 2(d) and (e). At $x = 0.01$ and 0.03, the $R_H$ is suddenly reduced at around $T = 150$ K, which arises from the formation of the SDW gap at the $E_F$ as reported previously in several non-doped compounds [22,23]. The strong temperature dependence is continuously reduced as $x$ is increased to $x \sim 0.3$, where the $R_H$ remains almost constant with $T$. For $x \geq 0.35$, although the $T$ dependence is still small, the $|R_H$ $(T)|$ begins to decrease below the specific temperature, $T_{RH}$, denoted by an arrow, and finally, the sign of $R_H$ is changed to positive in the low-temperature region at $x = 0.45$ and 0.52 [Fig. 2(d)].

In Fig. 3 (a) we plot the $T_{anom.1}$, $T_{anom.2}$, $T_c$, $T_{CO}$, and $T_{RH}$ of LaFeAsO$_{1-x}$H$_x$ against $x$. Besides these plots, the absolute value of the $\rho_{xx}$ $(T)$ is also shown on the contour plot. The $\rho_{xx}$ $(T)$ has two peaks over the whole temperature range at $x \sim 0$ and 0.5. After the SDW is suppressed at $x \geq 0.05$, the low resistive region spreads around $x \sim 0.2$ where the $T_{CO}$ is a maximum value while $T_c$ is a minimum. To analyze the normal conducting regime, the curve-fitting using the equation $\rho_{xx}$ $(T) = \rho_{xx}$ $(0) + AT^n$ was conducted in the two temperature regions of $50 \leq T \leq 100$ and $50 \leq T \leq 150$. The $x$ dependences of the exponent values of $n$ are plotted in Fig. 3(b). In the range of $0.07 \leq x \leq 0.30$, the $n$ is ~2, indicating that these three samples behave as a standard Fermi liquid. As $x$ increases to $x > 0.30$, however, the $n$ deviates from 2 and is sharply reduced to 1.18 at $x = 0.37$, and 0.54 at $x = 0.41$ (See the Supplemental Material SM4 [16]). Simultaneously, the $T_{CO}$ starts to decrease with $x$, and at $x = 0.41$, the $T_{CO}$ is unobservable, most likely to be below the $T_c$.

These non-monotonic changes of the $\rho_{xx}$ $(T)$ in each sample can be understood as the coherence-incoherence crossover, which was first proposed in $Ln$FeAsO$_{1-x}$F$_x$ [8]. At high temperatures, electrons are strongly scattered by the sizable magnetic moment of iron with a long- and short-time spin correlation function (incoherent region) [24,25]. As the temperature is decreased, however, the magnetic moments with a long-time spin correlation, which are sometimes referred to as frozen spins, are well screened by conducting electrons (coherent region) as seen in a Kondo lattice. Hence, the resistivity shows a drop at the crossover temperature, called the coherence scale $T^*$, below which the Fermi liquid is formed [8,25]. In this context, the spin-spin correlation is mainly driven not by the intra-orbital but by the inter-orbital interaction that the Hund's rule coupling is effective [26]. In the case of LaFeAsO$_{1-x}$H$_x$, the low resistive Fermi liquid state with $n \sim 2$ is only observed below the $T_{CO}$ around $x = 0.2$. After suppression of the $T_{CO}$ at $x = 0.41$, the highly resistive non-Fermi liquid state with a convex-shaped $\rho_{xx}$ $(T)$ curve appears. Theoretically, such a shape of $\rho_{xx}$ $(T)$ curve is reproduced by Haule *et al.* at the finite value of Hund's rule coupling $J_H$ [8], *i. e.*, the smaller $J_H$ stabilizes the Fermi liquid with the concave-shaped $\rho_{xx}$ $(T)$ curve below the coherence scale $T^*$, on the other hand, the larger $J_H$ enhances the absolute value of resistivity up to ~3 m$\Omega$cm at 300 K and reduces the $T^*$ above which the convex-shaped $\rho_{xx}$ $(T)$ curve appears. Based on the calculated results, the $T_{CO}$ that we define as inflection point on the $\rho_{xx}$ $(T)$ curve should be close to the $T^*$ since the slope of $d\rho/dT$ curve is expected to change from positive to negative as the temperature increased. In addition, the square-root relationship of $n \sim 0.5$ at $x = 0.41$ is known to be characteristic of the incoherent region and is experimentally observed in several compounds that have been investigated as candidates of the orbital-selective Mott phase (OSMP) for example, FeTe$_{1-x}$Se$_x$, SrRuO$_3$, and Sr$_2$RuO$_4$ [25,27–31]. Therefore, we suggest that the $T_{CO}$ is assigned to the coherence scale $T^*$.

As the $T_{CO}$ is strongly reduced for $x \geq 0.35$, another anomaly, abbreviated as $T_{RH}$, is observed in the $R_H$ ($T$) curve. In the two-band model including one electron-type with the density $n_e$ and mobility $\mu_e$ and one hole-type with the density $n_h$ and mobility $\mu_h$, the $R_H$ is expressed as [32]

$$R_H = \frac{1}{e} \frac{(n_e \mu_e^2 - n_h \mu_h^2) n_e \mu_e^2 + \mu_e^2 \mu_h^2 (n_e - n_h)(\mu_0 H)^2}{(n_e \mu_e + n_h \mu_h)^2 + \mu_e^2 \mu_h^2 (n_e - n_h)^2 (\mu_0 H)^2}. \qquad (1)$$

According to the Eq.(1) the field dependence of $\rho_{xy}$ ($H$) will become nonlinear. However, in the high field and temperature region except for the below the $T_{anom.1}$ at $x = 0.01$, 0.03, 0.52, and 0.58, the non-linear component is negligibly small (See the Supplemental Material SM1 [16]). Thus, the high field limit condition is applicable for the $R_H$, and the expression is reduced to

$$\lim_{H \to \infty} R_H = \frac{1}{e} \frac{1}{(n_e - n_h)}. \qquad (2)$$

In Fig. 3(c), we plot the $x$ dependence of the effective carrier number calculated from the equation $N_{eff.} = V/(2eR_H) = N_e - N_h$, where $N_e$ ($N_h$) is the number of electrons (holes) per Fe, and $V$ is the volume of unit cell. Except for the region of $0.30 \leq x \leq 0.58$, the variation of $N_{eff.}$ follows the simple relationship $N_{eff.} = -x$, indicating that electrons are the dominant carriers ($N_e \gg N_h$) in LaFeAsO$_{1-x}$H$_x$ [33]. However, for $x \geq 0.30$, the $N_{eff.}$ ($x$) curve begins to deviate from this relation, and shows an unrealistic positive value at $x = 0.45$ and 0.52. Considering the two facts that (*i*) for $x \geq 0.30$ the $T_{CO}$ is highly reduced, and (*ii*) for $x = 0.45$ and 0.52 their $T_{NS}$ coincide well with $T_{RH}$s, the singular behavior in $R_H$ ($T$) is ascribed to the strong scattering of electrons by the frozen spins above the $T_{CO}$ and ordered AFM spins below the $T_{anom.2}$ [34].

Here, we briefly summarize the transport and magnetic properties of LaFeAsO$_{1-x}$H$_x$

with $T$ and $x$. In the low $x$ region, after suppression of the SDW at $x = 0.01$, the coherent Fermi liquid state develops below the high $T_{CO}$ around $x = 0.2$. As $x$ increases to $x \geq 0.3$, the $T_{CO}$ decreases, above which the non-Fermi liquid states, characterized by the reduced exponent, appear (incoherent region). The heavy doping above $x \geq 0.45$ induces the AF2 ordering and the sign change of $R_H$ below the $T_{anom.2}$, where the resistivity exhibits a second maximum around $x \sim 0.5$.

It should be noted that the mechanism for the AF2 is not the SDW. There are two results that support this interpretation. One is the non-nested Fermi surface of LaFeAsO$_{0.5}$H$_{0.5}$. Figures 3(d) and (e) show the Fermi surfaces calculated for tetragonal LaFeAsO and LaFeAsO$_{0.5}$H$_{0.5}$, respectively. In these figures, the black arrows indicate the two-dimensional propagation vector, $\boldsymbol{Q} = (\pi/2, \pi/2)$, for each AFM ordering [12,35]. The Fermi surface of LaFeAsO has a clear nesting vector along with the $\boldsymbol{Q}$, connecting between the hole and electron cylinders at the Γ and M points, respectively. In contrast, the Fermi surface of LaFeAsO$_{0.5}$H$_{0.5}$ has no nesting in this direction because the electron surfaces are highly expanded by the electron doping. Another result is the temperature dependence of the $R_H$. In the non-doped sample, we observed a large enhancement of the $|R_H|$ below the $T_N$ owing to the reduction of conduction electron by opening of the SDW gap, whereas at $x = 0.52$, the $|R_H|$ did not vary much above and below the $T_N$. Thus, the AF2 is not described by the itinerant picture, but reliably by the more localized magnetism. The large magnetic moment found in the AF2 also supports this explanation. This inconsistency between the $\boldsymbol{Q}$ and nesting vectors, and the absence of the SDW-gap, are also found in FeTe [36].

The last problem for discussion is the reason why the electron doping increases the electron correlation and reduces the coherence scale $T_{CO}$. So far, the crossover phenomenon from a metal into a bad metal or insulator on heating has been observed in several compounds. Indeed, recently, Yi *et al.* directly observed a loss of the quasiparticle spectral weight of the 3$d_{xy}$ orbital above the specific temperature, $T_{OSMP}$, in FeTe$_{0.56}$Se$_{0.44}$, K$_{0.76}$Fe$_{1.72}$Se$_2$, and mono-layer FeSe by using ARPES [37]. A common feature of these compounds is that the less-occupied Fe $3d_{xy}$ orbital is responsible for the electronic correlation. Hence, as for the effect of doping, it is the consensus that hole doping enhances the electron correlation as the occupation approaches 1, while electron doping is prone to reduce the correlation. To understand this observation, we examine the effect of electron doping on the electronic structure of LaFeAsO$_{1-x}$H$_x$ in Sec. IV-B and C.

## B. Band structures of LaFeAsO weighted with molecular orbitals

Before we consider the electronic states of LaFeAsO$_{1-x}$H$_x$, we wish to discuss another problem in the field of iron-based superconductors, that is, the two-Fe versus one-Fe description, which is important but has rarely been discussed to date [38,39]. The crystallographic unit cell of iron-based superconductors has two irons tetrahedrally coordinated by arsenic. Each $3d$-orbital makes bonding- and anti-bonding-molecular orbitals composed of in-phase or out-of-phase linear combinations of the atomic orbitals. Therefore, there are ten $3d$-bands in the first Brillouin zone (BZ). This is the two-Fe description. On the other hand, many theorists use the one-Fe description by unfolding the half of bands inside the two-Fe BZ to the larger one-Fe BZ. This treatment can reduce the number of bands to five and allow us simply to construct the physical model, such as the Hubbard model. Herein, we adopt the former case, the two-Fe description.

Figure 4 (a) shows a $3d_{xy}$-molecular orbital of Fe at the Γ point. The phase of $3d_{xy}$ at the Fe1 = (3/4, 1/4, 1/2) site is inversed at the Fe2 = (1/4, 3/4, 1/2) site, so we name such a configuration as out-of-phase. In Fig. 4(b), the doubly-degenerate orbital configurations of As-$4p_Z$ at the M point are depicted. Note that the orbital configuration shown on the right side is in-phase because the phase of the $4p_Z$ orbital at the As1 = (1/4, 1/4, −Z) site is the same as that at the As2 = (3/4, 3/4, Z) site, as illustrated in Fig. 4(c). The phase of the orbital is also changed by the translation of the unit cell if the state is located at without Γ point. For instance, in the case of the M point, shown in Fig. 4(b), the phase of the orbital is inversed by translation along the **a**- and **b**-axes and is unchanged by translation along the **c**-axis. Owing to the difference of the site symmetry, we use two coordinates for each atomic orbital; for the cases of Fe and O, we adopt the $x$ and $y$ axes along with the directions of nearest neighbor Fe-Fe, while for As, the $X$ and $Y$ axes along with **a**- and **b**-axes of the tetragonal unit cell are used.

To construct the molecular orbital in LaFeAsO$_{1-x}$H$_x$, we introduce an MLWF calculated from DFT band structures [20], and define a new function extracting the information of the phase of each MLWF. According to Ref. [40], the $n$th Bloch state calculated by a Slater–Koster interpolation using the MLWF is expressed as

$$|u_{nk}^{(H)}> = \sum_m^M |u_{mk}^{(W)}> U_{mn}(\boldsymbol{k}),\tag{3}$$

where $n$, $M$, and $U_{mn}$ are the band-index, number of MLWF, and element of the $M \times M$ unitary matrix, respectively. The quantity with a superscript (W) or (H) is referred to as belonging to the Wannier or Hamiltonian gauge, respectively. Each component of the right side of Eq. (3) is originally complex. Therefore, if one calculates the inner product between any two MLWFs, that is, $m = \alpha$ and $\beta$ on the right side of Eq. (3), the sign of the real part of the product (positive or negative) reflects the relative phase between the $\alpha$ and $\beta$th MLWFs (in-phase or out-of-phase). Thus, we define such a quantity $F_n^{\alpha, \beta}(\boldsymbol{k})$ as

$$F_n^{\alpha,\beta}(\boldsymbol{k}) = 2\mathrm{Re}[U_{\alpha n}^{\dagger}(\boldsymbol{k}) < u_{\alpha\boldsymbol{k}}^{(W)}||u_{\beta\boldsymbol{k}}^{(W)}> U_{\beta n}(\boldsymbol{k})].\tag{5}$$

The calculated electronic structures of LaFeAsO are shown in Fig. 5. The Fe-3$d$ bands are across the $E_F$ from $-2$ to 2 eV, and As-4$p$ and O-2$p$ bands are located below them [Fig. 5 (a)]. Figure 5 (b) and (c) show band structures of LaFeAsO colored using the $F_n^{\alpha, \beta}(\boldsymbol{k})$ values with $\alpha, \beta = 3d_{xy}$ at the Fe1- and Fe2-site, and with $\alpha, \beta = 4p_z$ at the As1-and As2-site, respectively. In these figures, the thickness of the colored band represents the contribution from each atomic orbital, and the red and blue colors indicate the contributions from the in-phase and out-of-phase molecular orbitals, respectively. Figure 5(b) shows that the in-phase Fe-3$d_{xy}$ band (Fe-3$d_{xy}^{\mathrm{in}}$) lies in the $E_F$ over the whole $\boldsymbol{k}$-space, while the out-of-phase Fe-3$d_{xy}$ band (Fe-3$d_{xy}^{\mathrm{out}}$) is highly dispersed from $-4$ to 1 eV. At the M point, these two bands are degenerate at $E - E_F = -0.4$ eV. As for arsenic, the in-phase As-$4p_Z$ (As-$4p_z^{in}$) is populated above $E_F$, while the out-of-phase As-$4p_Z$ (As-$4p_z^{out}$) is located at lower energy of $E - E_F = -4$ eV. These destabilizations of Fe-$3d_{xy}^{out}$ and As-$4p_z^{in}$ around the Γ point originate from the formation of an anti-bonding state between them. In contrast, the Fe-$3d_{xy}^{in}$, colored mainly in red in Fig. 5(b), shows no hybridization with As-$4p_z^{in}$ and thus, has a very narrow bandwidth. The difference in the degree of hybridization results from the symmetry of each molecular orbital around the Γ point. Table 1 summarizes the symbols for the irreducible representations (Irrep.) of each molecular orbital at the Z and Γ points. In principle, molecular orbitals with the same Irrep. can hybridize each other, and the resulting states form new bonding and anti-bonding states between inter-molecular orbitals. According to Table 1, it is shown that the Fe-$3d_{xy}^{in}$ shares its Irrep. with O-$2p_z^{out}$ at the Γ point ($Γ_{3+}$), and H-$1s^{out}$ at the Z point ($Z_{4-}$), whereas the Fe-$3d_{xy}^{out}$ shares with As-$4p_z^{in}$ at both the Z ($Z_{1+}$) and Γ ($Γ_{2-}$) points as well as with the O-$2p_z^{in}$ and H-$1s^{in}$ at the Z or Γ point. Figure 6(a) and (b) show the enlarged band structures of Fig. 5(b) and (c) in the Z–Γ line, respectively. In Fig. 6(a), the symbols of Irrep.s are also indicated at the Z and Γ points. As shown in the Table 1, each molecular orbital composed of Fe-$3d_{xy}$ or As-$4p_z$ is distributed in the band with the corresponding Irrep., and some of them co-exist on the band with the common Irrep., that is, both Fe-$3d_{xy}^{out}$ and As-$4p_z^{in}$ orbitals are found in the $Z_{1+}$ at the Z point and $Γ_{2-}$ at the Γ point. However, the contribution from O-$2p_z$, shown in Fig. 6(c), is highly localized below 4 eV, even though it can be mixed with As-$4p_z^{out}$ at the Z point and with Fe-$3d_{xy}^{in}$ at the Γ point. This is because the O-$2p_z$ is energetically and spatially far from those of Fe-$3d_{xy}$ and As-$4p_Z$ [41]. Figure 6(d–f) illustrate the relevant molecular orbital configurations: a bonding state between Fe-$3d_{xy}^{out}$ and As-$4p_z^{in}$ [Fig. 6(d)], an anti-bonding state between Fe-$3d_{xy}^{out}$ and As-$4p_z^{in}$ [Fig. 6(e)], and a nonbonding state of Fe-$3d_{xy}^{in}$ [Fig. 6(e)]. Hereafter, we shall call the representative three orbital configurations *dp*, *dp\**, and *dd\**, respectively.

It is worth emphasizing that these three states *dp*, *dp\**, and *dd\** are averaged within a single $3d_{xy}$ orbital in the one-Fe description. By taking care of the two-dimensional unit cell of the one-Fe description that is reduced to a $1/\sqrt{2} \times 1/\sqrt{2}$ cell and rotated by 45 °, one can find the relation between the orbital configuration of $3d_{xy}$ in the two- and one-Fe descriptions as follows: (*i*) the Fe-$3d_{xy}^{\text{in}}$ at the Γ point (*dd\**) is identical to the portion of the $3d_{xy}$ band at the Γ point for the one-Fe BZ (Γ'); (*ii*) the Fe-$3d_{xy}^{\text{out}}$ at the Γ point [See Fig. 4(a)] is transferred to the portion of the $3d_{xy}$ band at the M' point, and splits into the *dp* and *dp\** states by hybridizing with As-$4p_z^{\text{in}}$, the Fe-$3d_{xy}^{\text{out}}$. Therefore, the *dd\**, *dp*, and *dp\** coexist in the single $3d_{xy}$ band in the one-Fe description. Such a correspondence is seen in Fig. 7 of Ref. [42].

### C. Effects of electron doping on the electronic structure of LaFeAsO$_{1-x}$H$_x$

Next, we examine the effect of electron doping. Figure 7(a) and (d) show the DOS of LaFeAsO$_{0.5}$H$_{0.5}$ and LaFeAsO$_{0.25}$H$_{0.75}$, respectively. Compared with the DOS of LaFeAsO, the band width of Fe becomes narrower from ~4 eV at $x = 0$ to ~3 eV at $x = 0.5$ and 0.75. The H $1s$ band is located below the Fe-band. The $F_n^{\alpha, \beta}(\boldsymbol{k})$ weighted band structures of LaFeAsO$_{0.5}$H$_{0.5}$ are shown in Fig.7(b) and 7(c). Most of the doped electrons occupy the *dp\** state, which is stabilized to below −1.0 eV. However, the *dd\** state remains almost unchanged at any $\boldsymbol{k}$-point by the doping or, rather, is shifted upwards as seen at the Γ point. This up-shift of the $3d_{xy}$ band around the Γ point has been reported previously in the case of using a fluorine dopant [13,43]. A similar variance by electron doping is also observed in LaFeAsO$_{0.25}$H$_{0.75}$ [Fig. 7(e) and (f)].

To calculate the electron occupation in Fe-$3d_{xy}^{\text{in}}$ and Fe-$3d_{xy}^{\text{out}}$ against $x$, the BZ-integrated $F_n^{\alpha, \beta}(\boldsymbol{k})$ are shown in Fig. 8(a–c). At $x = 0$, the narrow *dd\** state is located just below the $E_F$, and the Fe-$3d_{xy}^{\text{out}}$ shows two peaks at $E = 1$ and −3 eV, which are anti-bonding (*dp\**) and bonding (*dp*) states with As-$4p_z^{\text{in}}$, respectively. As $x$ increases to 0.5 or 0.75, the energy of the *dp\** state decreases substantially, whereas the energy of the *dd\** state increases. Consequently, the *dd\** state approaches half-filled state at $x = 0.75$. It is reported that the implementation of dynamical mean field theory tends to reduce the occupation of $3d_{xy}$ compared with the conventional GGA results. Hence, the occupation of the $dd*$ should decrease even more if the electron correlation is properly treated [10,44]. Figure 8 (d) summarizes the effect of electron doping using a molecular orbital diagram at the $\Gamma$ point. The left side of the diagram shows the energy levels of Fe-$3d_{xy}{}^{\text{in}}$ and Fe-$3d_{xy}{}^{\text{out}}$, and those of As-$4p_Z{}^{\text{in}}$ and As-$4p_Z{}^{\text{out}}$ are on the right side. Because the Fe-$3d_{xy}{}^{\text{out}}$ and As-$4p_Z{}^{\text{out}}$ are bonding states, they are located at lower energy. Including the bond formation between orbitals of Fe and As, the energy level of Fe-$3d_{xy}{}^{\text{out}}$ splits into the bonding ($dp$) and anti-bonding ($dp*$) levels, while the Fe-$3d_{xy}{}^{\text{in}}$ is unchanged and behaves as the non-bonding state ($dd*$). When electrons are doped in this system, these electrons occupy the $dp*$ state and work to weaken the Fe-As bond owing to the anti-bonding nature of $dp*$. Finally, the reduced energy of the $dp*$ state approaches the initial energy level of $3d_{xy}{}^{\text{out}}$.

Such doping effects are also observed in the resultant changes of the structural parameters. The $x$ dependence of the Fe-Fe distance ($d_{\text{Fe}}$), Fe-As distance ($d_{\text{Fe-As}}$), and the height of the As from Fe plane ($h_{\text{As}}$) are plotted in Fig. 8(e). As $x$ increases, the $d_{\text{Fe-As}}$ and $h_{\text{As}}$ gradually become longer, while the $d_{\text{Fe}}$ decreases, which results from the anti-bonding configuration of the Fe-$3d_{xy}$-As-$4p_Z$ orbitals and bonding configuration of the Fe-$3d_{xy}$-Fe-$3d_{xy}$ orbitals in the $dp*$ state, respectively. In Fig. 8(f), we show the calculated hopping parameters of the $3d_{xy}$ orbital to the nearest ($t_1$, filled diamond) and next-nearest-neighbor Fe atom ($t_2$, open diamond) obtained from the $d$ model containing only ten $3d$ orbitals. Owing to the large $h_{\text{As}}$ at $x = 0.5$ and $0.75$, the indirect hopping through As is reduced, and then, the effective $t_1$ is almost cancelled by the direct hopping with the opposite sign in the heavily doped region [10, 43].

Here, we discuss again the phase diagram of LaFeAsO$_{1-x}$H$_x$ based on the results of our calculations. In the low $x$ region, the Fe-$3d_{xy}{}^{\text{in}}$ band is located below the $E_{\text{F}}$, and the other bands, for example, Fe-$3d_{xy}{}^{\text{out}}$ and Fe-$3d_{yz/zx}$, are largely dispersed. Hence, the local spins are well screened by the conduction electrons and the coherence scale, denoted by $T_{\text{CO}}$ below which the Fermi liquid is formed, will be enhanced. This is the case for $0 \leq x \leq 0.2$, except for the accidental SDW phase at $x \sim 0$. As $x$ is increased, although the total bandwidth of Fe is still large, the narrow Fe-$3d_{xy}^{in}$ band approaches the half-filled state, and the $t_1$ and $t_2$ of the $3d_{xy}$ are reduced. Experimentally, the coherence scale is sharply decreased for $x > 0.2$, and after complete suppression of the $T_{CO}$ at $x = 0.41$, the non-Fermi liquid states, characterized by the $\sqrt{T}$-dependency of resistivity, appear. According to the Kondo problem of a half-filled $3d$ or $4f$ magnetic impurity embedded in a metallic sea, the electrons occupying the half-filled orbital cannot exchange the orbital moment with the conduction electron owing to the Hund's rule coupling [45,46]. Thus, in this case, the coherence scale (or the Kondo temperature) is reduced exponentially because the screening to the local spins in the half-filled orbital is highly restricted. For $x > 0.45$, a long-range magnetic ordering (AF2) takes place below $T \sim 100$ K. This is also reasonable because the magnetic moments inside such an incoherent region carrying an extremely large entropy tend to order or be quenched to spin glass at low temperature [11]. Although the magnetic structure of AF2 is determined as a commensurate stripe-type ordering, the muon spin relaxation results in a significant damping in contrast to a clear precession in AF1, suggesting an inhomogeneity in the local magnetic fields [12]. The heavy doping above $x > 0.58$ seems to weaken the electronic correlation. At $x = 0.66$, the absolute value of the resistivity is reduced, and the kink at $T_{anom.1}$ and $T_{anom.2}$ becomes obscure. Moreover, the singular behavior in the $R_H (T)$ curve is also diminished. These results would suggest the over-doping of electrons into the Fe-$3d_{xy}^{in}$.

The transport properties of the LaFeAsO$_{1-x}$H$_x$ and the other electron-doped iron pnictides, for example, Ba(Fe$_{1-x}$Co$_x$)$_2$As$_2$, are in sharp contrast. Although the phase diagram of the Ba(Fe$_{1-x}$Co$_x$)$_2$As$_2$ is very similar to that of LaFeAsO$_{1-x}$H$_x$ for $0 \leq x \leq 0.2$, in the heavily doped region above $x \geq 0.2$, only the Fermi liquid state is attained [14]. Such a difference in the heavily doped region between the 1111 and 122 types can also be observed from the emergence of spin fluctuations and superconductivity in LaFeAsO$_{1-x}$H$_x$, which are absent in Ba(Fe$_{1-x}$Co$_x$)$_2$As$_2$ [13,47–49].

Finally, we would like to discuss the boundary between the spin-freezing ($0.35 \leq x \leq 0.41$) and spin-ordering ($0.45 \leq x \leq 0.66$) regions, which is characterized by the appearance of $T_{RH}$. To the best of our knowledge, the sudden decrease of the $|R_H|$ or the following sign change is observed in a few compounds, FeTe$_{1-x}$Se$_x$, (K or Rb)$_y$Fe$_{2-\delta}$Se$_2$, and over-doped Ba$_{1-z}$K$_z$Fe$_2$As$_2$ with $z \geq 0.45$ [50–55]. As for FeTe$_{1-x}$Se$_x$, the $T_{RH}$ may be assigned to the $T_{OSMP}$ above which the $3d_{xy}$ orbital loses all spectral weight while the other orbitals remain itinerant, because the $T_{OSMP}$ of 101 K at $x = 0.44$ is very close to the $T_{RH}$ ~ 100 K at $x = 0.45$ [36, 52]. Curiously, similar to the phase diagram of LaFeAsO$_{1-x}$H$_x$, the $T_{RH}(x)$ curve of FeTe$_{1-x}$Se$_x$ touches the $T_N$ in $0.00 \leq x \leq 0.08$, and diverges above the optimum $T_c$ at $x = 0.45$ if we take the $T_{RH}(x)$ from Fig. 2(b) in Ref. [51] for the data at $0.00 \leq x \leq 0.08$ and Fig. 4(b) in Ref. [52] for the data at $0.14 \leq x \leq 0.45$. Hence, we believe that the $T_{RH}(x)$ of LaFeAsO$_{1-x}$H$_x$ traces the $T_{OSMP}(x)$.

# VI. CONCLUSION

We systematically examined the transport properties and band structures in the normal conducting states of electron-doped LaFeAsO$_{1-x}$H$_x$ over a wide range of $x$. The main results of this paper are summarized as follows;

(*i*) From the resistivity measurements, we detected a clear crossover phenomenon from the coherent Fermi liquid around $x = 0.2$ to the incoherent non-Fermi liquid around $x = 0.4$, which accompany the substantial depression of the coherent scale $T_{CO}$ with $x$. The $\sqrt{T}$ dependency of resistivity at $x = 0.41$ suggested that this sample is in the incoherent region.

(*ii*) In the boundary between the SC2 and AF2 in the range of $0.35 \leq x \leq 0.41$, the strong magnetic scattering forced the $|R_H|$ to be reduced. The subsequent sign-change of $R_H$ without opening of the SDW gap for $0.45 \leq x \leq 0.58$ as well as the calculated non-nested Fermi surface at $x = 0.5$ indicate the more localized nature of AF2 as compared to AF1 phase at non-doped sample.

(*iii*) By employing the concept of the molecular orbital, we revealed that the narrow anti-bonding $3d_{xy}$ band (in-phase) approaches the half-filled regime as $x$ is increased.

(*iv*) Considering the finite size of the Hund's rule coupling, the substantial depression of the coherent scale with $x$ were reasonably explained by the increased effective Coulomb repulsion $U_{eff}$ in the narrow anti-bonding $3d_{xy}$ band approaching the half-filled regime. The following transition into AF2 was understood as the quenching or ordering of less-screened spins with a large entropy at low temperature. The large magnetic moment was interpreted as a result of the enhanced $U_{eff}$ and the reduced hopping $t$ in the $3d_{xy}$ orbital.

Comparing these results with those of the conventional electron-doped iron pnictides, for example, $Ba(Fe_{1-x}Co_x)_2As_2$, we indicate that the normal and superconducting properties of the $LaFeAsO_{1-x}H_x$ should be recognized to be strongly influenced by the local spins ordered as the AF2 phase, in other words, the AF2 is the parent phase of $LaFeAsO_{1-x}H_x$, which may be responsible for the high-$T_c$ superconductivity in the 1111 type.

## VII. ACKNOWLEDGEMENT

We would like to thank H. Fukuyama, K. Terakura, S. Ishibashi, H. Seo, and Z. P. Yin for valuable discussions. This study was supported by the MEXT Element Strategy Initiative Project to form a research core and Grant-in-Aid for Young Scientists (B) (Grant No. 26800182) from JSPS.

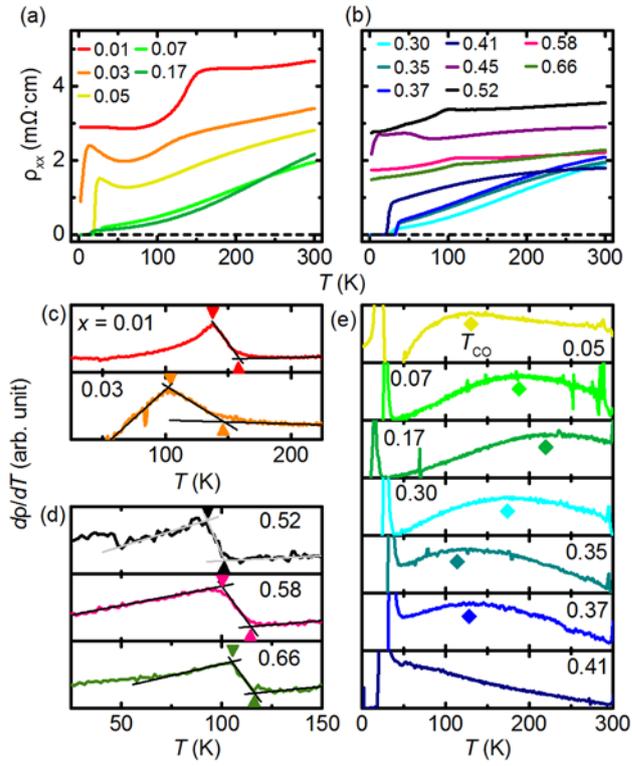

FIG. 1. Color Temperature dependence of resistivity, $\rho_{xx}$, in LaFeAsO$_{1-x}$H$_x$ (a,b) in the range of $0.01 \leq x \leq 0.17$ (a) and $0.30 \leq x \leq 0.66$ (b). (c–e) The temperature differentials of $\rho_{xx}$. The triangle, inverse triangle, and diamond denote the $T_{anom.1}$, $T_{anom.2}$, and $T_{CO}$, respectively. The $T_{CO}$ are determined at the temperature that the second derivatives of $\rho_{xx}$ is zero. To obtain the differential of the $d\rho_{xx}(T)/dT$, we used the Savitzky–Golay scheme.

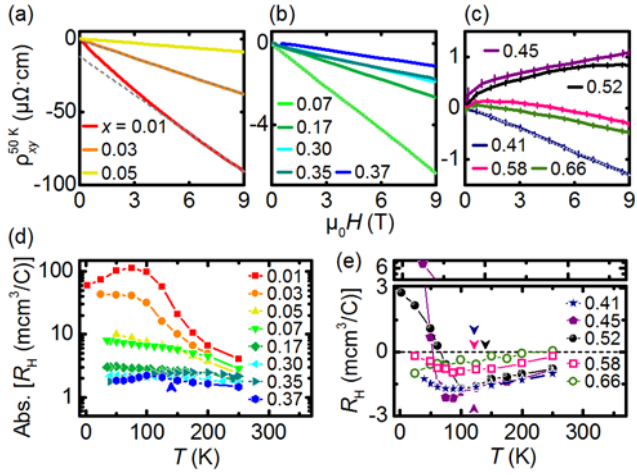

FIG. 2. Color Hall effects in LaFeAsO$_{1-x}$H$_x$. (a–c) The field dependence of $\rho_{xy}$ at $T = 50$ K in the range of $0.01 \leq x \leq 0.05$ (a), $0.07 \leq x \leq 0.37$ (b), and $0.41 \leq x \leq 0.66$ (c). The straight dashed line is included as a guide for the eyes. (d) The temperature dependence of the absolute value of $R_H$ in the range of $0.01 \leq x \leq 0.37$. (e) The temperature dependence of $R_H$ in the range of $0.41 \leq x \leq 0.66$. The arrows mark the temperature ($T_{RH}$) at which the $R_H$ ($T$) curves approach the positive side as the $T$ decreased.

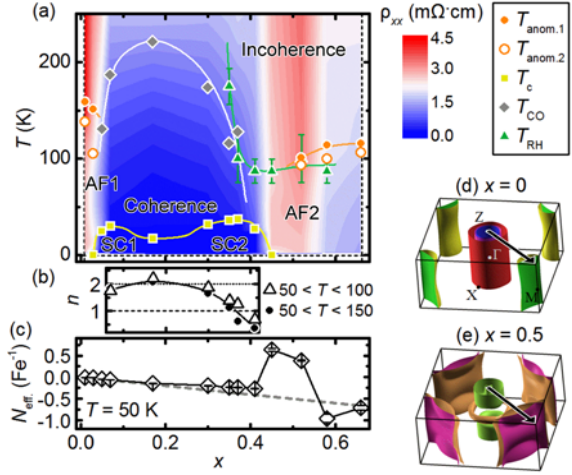

FIG. 3. Color (a) Phase diagram of LaFeAsO$_{1-x}$H$_x$. The yellow squares, orange filled circles, orange open circles, green triangles, and gray diamonds show the onset $T_c$, $T_{anom.1}$, $T_{anom.2}$, $T_{RH}$, and $T_{CO}$, respectively. The $T_c$, $T_{anom.1}$, $T_{anom.2}$, and $T_{CO}$ are determined from resistivity curves $\rho_{xx}$ $(T)$, whereas the $T_{RH}$ is determined from $R_H$ $(T)$ curves. (b) The $x$ dependency of $n$ in the superconducting samples with $x = 0.07$–$0.41$. The exponent $n$ is derived from the fitting using the equation $\rho_{xx}$ $(T) = \rho_0 + AT^n$. (c) The $x$ dependency of the number of effective carrier per Fe is defined as $N_{eff.} = N_e - N_h$, where the $N_e$ and $N_h$ are the number of electrons and holes per Fe, respectively. The straight dashed line indicates the value calculated by $N_{eff.} = -x$. (d, e) Fermi surfaces of LaFeAsO (d) and LaFeAsO$_{0.5}$H$_{0.5}$ (e). The black arrows inside the BZ indicate the 2-dimensional propagation vector $\boldsymbol{Q}$ observed in their AFM structures: $\boldsymbol{Q} = (\pi/2, \pi/2)$ [35, 12].

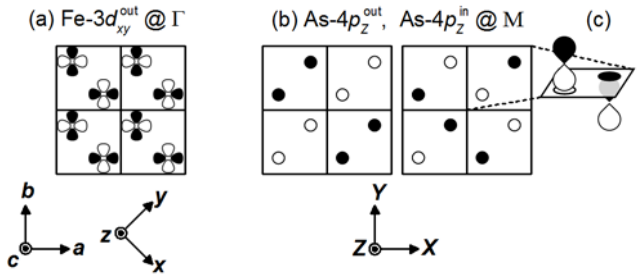

FIG. 4. (a) Out-of-phase-Fe-$3d_{xy}$ molecular orbital (Fe-$3d_{xy}^{\text{out}}$) in 2×2 unit cells at the $\Gamma$ point. (b) The two configurations of the As-$4p_Z$ molecular orbitals in 2×2 unit cells at the M point. The left is out-of-phase and right is in-phase. Here, the filled and open circles are the lobes of the $4p_Z$ orbitals touching the Fe-plane, as sketched in Fig. 4(c). At the $\Gamma$ point, the phase of the molecular orbital within the unit cell is unchanged by any translation, whereas at the M point, the phase is changed by translation along the $\boldsymbol{a}$- and $\boldsymbol{b}$-axes and is unchanged by translation along the $\boldsymbol{c}$-axis. We choose the $x$ and $y$ axes in the Fe-Fe direction, while the $X$ and $Y$ axes are along with $\boldsymbol{a}$- and $\boldsymbol{b}$-axes of the tetragonal unit cell. The coordinates of the atoms inside the unit cell are (3/4, 1/4, 1/2) and (1/4, 3/4, 1/2) for Fe1 and Fe2, (1/4, 1/4, −$Z$) and (3/4, 3/4, $Z$) for As1 and As2, and (3/4, 1/4, 0) and (1/4, 3/4, 0) for O1 and O2.

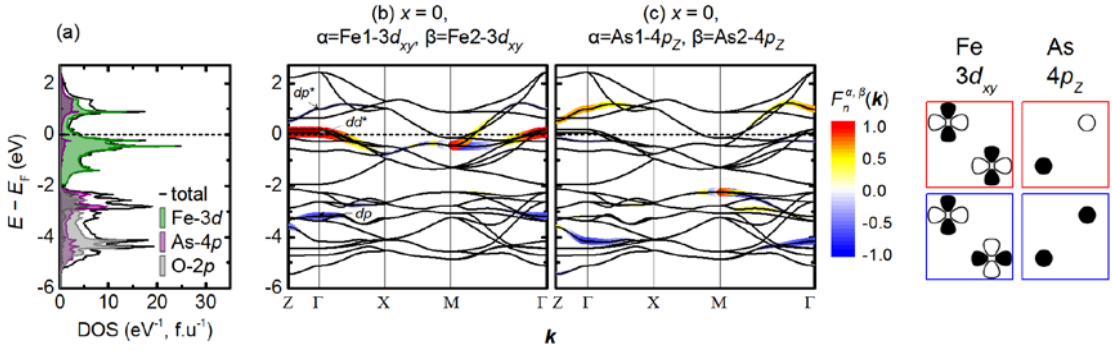

FIG. 5. Color (a) Total density of states (black solid line) and partial DOS of Fe-3$d$ (green), As-4$p$ (purple), and O-2$p$ (gray) calculated using the Slater–Koster-interpolation based on the Wannier orbital basis. (b, c) The band structure colored by $F_n^{\alpha, \beta}(\boldsymbol{k})$ with $\alpha = 3d_{xy}$ at the Fe1 site and $\beta = 3d_{xy}$ at the Fe2 site(b), and by $F_n^{\alpha, \beta}(\boldsymbol{k})$ with $\alpha = 4p_z$ at the As1 site and $\beta = 4p_z$ at the As2 site (c). The color indicates the sign of $F_n^{\alpha, \beta}$, and the width of the colored fat-band is proportional to the contribution of the 3$d_{xy}$ (b) and 4$p_Z$ orbitals (c). Orbital configurations inside the unit cell are depicted on the right side of the color bar; the red cells represent that the $F_n^{\alpha, \beta}(\boldsymbol{k})$ is positive (in-phase) while the blue represents that the $F_n^{\alpha, \beta}(\boldsymbol{k})$ is negative (out-of-phase).

TABLE 1. Symbols used to identify the irreducible representations of molecular orbitals in LaFeAsO$_{1-x}$H$_x$ with the space group $P4/nmm$. The irreducible representations (Irrep.) and the symbols are determined based on Ref. [56].

| | | Fe | | | | | | | | |
|---|---|---|---|---|---|---|---|---|---|---|
| | | $3d_{xy}$ | | $3d_{yz}$ | | $3d_{zx}$ | | $3d_{x2-y2}$ | | $3d_{z2}$ | |
| | | in | out | in | out | in | out | in | out | in | out |
| Z | | 4− | 1+ | 5− | 5+ | 5− | 5+ | 3− | 2+ | 2− | 3+ |
| Γ | | 3+ | 2− | 5+ | 5− | 5+ | 5− | 4+ | 1− | 1+ | 4− |

(*continued*)

| As | | | | | | O | | | | | | H | |
|---|---|---|---|---|---|---|---|---|---|---|---|---|---|
| $4p_X$ | | $4p_Y$ | | $4p_Z$ | | $2p_x$ | | $2p_y$ | | $2p_z$ | | $1s$ | |
| in | out | in | out | in | out | in | out | in | out | in | out | in | out |
| 5+ | 5− | 5+ | 5− | 1+ | 2− | 5− | 5+ | 5− | 5+ | 2− | 3+ | 1+ | 4− |
| 5− | 5+ | 5− | 5+ | 2− | 1+ | 5− | 5+ | 5− | 5+ | 2− | 3+ | 1+ | 4− |

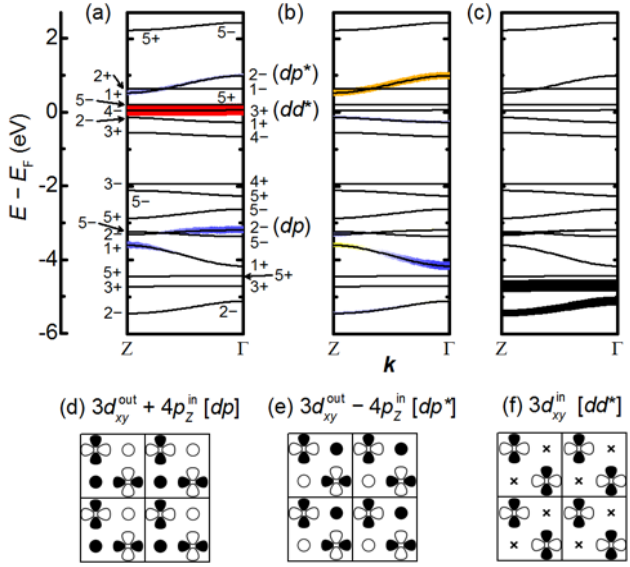

FIG. 6. Color (a, b) Enlarged band structures of LaFeAsO in the Z-Γ line of Fig. 5(b) for Fig. 6(a) and that in Fig. 5(c) for Fig. 6(b). In Fig. 6(a), the corresponding Irrep.s are also added in the bands at the Z and Γ points. (c) Contribution from the oxygen-2$p$ orbitals to the band structures of LaFeAsO in Z-Γ. (d-f) Orbital-configurations of some bands in 2×2 unit cells at the Γ point. Each orbital-configuration corresponds to a band marked by *dp*, *dp\**, and *dd\** in Fig. 6(a) in parentheses.

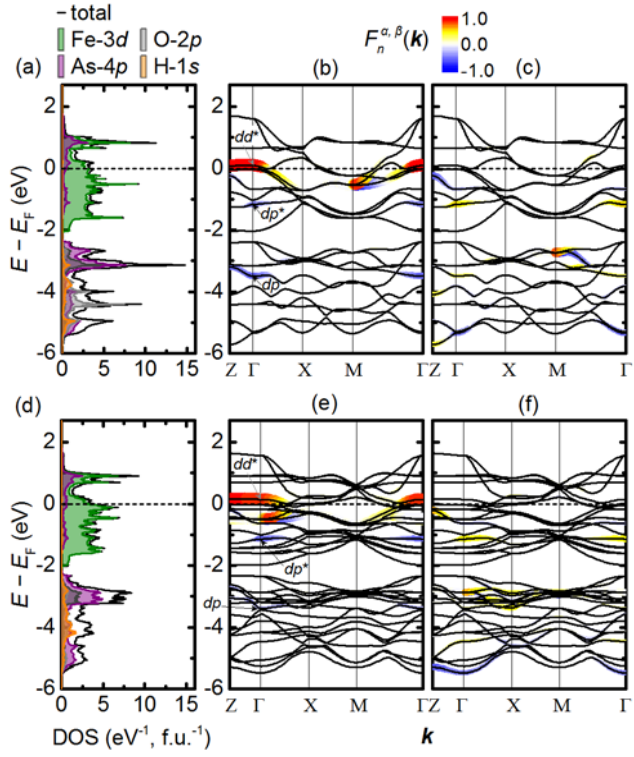

FIG. 7. Color (a, d) Total (black solid line) and partial DOS of Fe (green), As (purple), O (gray), and H (orange) at $x = 0.5$ (a) and 0.75 (d). (b, c) The band structures at $x = 0.5$ colored by $F_n^{\alpha, \beta}(\boldsymbol{k})$ with $\alpha = 3d_{xy}$ at the Fe1 site and $\beta = 3d_{xy}$ at the Fe2 site (b), and by $F_n^{\alpha, \beta}(\boldsymbol{k})$ with $\alpha = 4p_z$ at the As1 site and $\beta = 4p_z$ at the As2 site (c). (e, f) The band structures at $x = 0.75$ colored by $F_n^{\alpha, \beta}(\boldsymbol{k})$ with $\alpha = 3d_{xy}$ at the Fe1 site and $\beta = 3d_{xy}$ at the Fe2 site (e), and by $F_n^{\alpha, \beta}(\boldsymbol{k})$ with $\alpha = 4p_z$ at the As1 site and $\beta = 4p_z$ at the As2 site (f). Because we used the $\sqrt{2} \times \sqrt{2} \times 1$ supercell for the calculation, the BZ at $x = 0.75$ are folded from the primitive tetragonal BZ at ambient temperature.

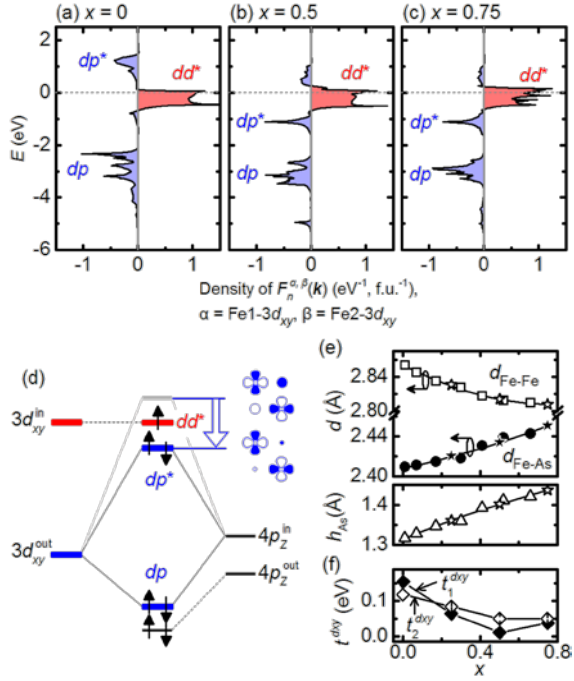

FIG. 8 Color (a-c) The energy dependence of BZ-integrated $F_n^{\alpha,\beta}(\boldsymbol{k})$ with $\alpha = 3d_{xy}$ at Fe1 site and $\beta = 3d_{xy}$ at Fe2 site at $x = 0$ (a), 0.5 (b), and 0.75 (c). (d) A schematic molecular-orbital-diagram of LaFeAsO$_{1-x}$H$_x$ at $\Gamma$ point constructed based on the $F_n^{\alpha,\beta}(\boldsymbol{k})$-weighted-band structures shown in Fig. 5 and 7. The left side of the diagram shows the energy levels of Fe-$3d_{xy}^{\mathrm{in}}$ (red) and Fe-$3d_{xy}^{\mathrm{out}}$ (blue), and those of As-$4pz^{\mathrm{in}}$ and As-$4pz^{\mathrm{out}}$ in the right side. Between them, the energy levels of bonding and antibonding states formed between inter-molecular orbitals are shown. The out-of-phase $4pz$ is stabilized by forming the bonding orbital with the in-phase-$3d_{z2}$ orbital of Fe and the $1s$ orbital of H shown as the dashed line (See Table 1). (e) The doping dependence of experimental Fe-Fe ($d_{\mathrm{Fe\text{-}Fe}}$, open square), Fe-As bond-lengths ($d_{\mathrm{Fe\text{-}As}}$, filled circle), and As-height ($h_{\mathrm{As}}$) from Fe-plane ($d_{\mathrm{Fe\text{-}As}}$, open triangle). These data were obtained by X-ray diffraction under room temperature. The solid lines are calculated by using the fitting curves of the $x$-variation of $a$-axis, $c$-axis, and fractional coordinates of La and As. The values marked by stars are obtained from the fitting curves and used for DFT calculations. (f) The doping dependence of calculated hopping parameters of $3d_{xy}$ orbital to the nearest ($t_1$, filled diamond) and next nearest neighbor Fe atom ($t_2$, open diamond) obtained from the $d$ model containing only ten $3d$ orbitals of Fe.

# Supplemental Materials For Heavy Electron doping-induced antiferromagnetic phase as the parent for iron-oxypnictide superconductor LaFeAsO$_{1-x}$H$_x$

Soshi Iimura[1], Satoru Matsuishi[2], and Hideo Hosono[1, 2]

[1] Laboratory for Materials and Structures, Tokyo Institute of Technology, Yokohama 226-8503, Japan

[2] Materials Research Center for Element Strategy, Tokyo Institute of Technology, Yokohama 226-8503, Japan

Correspondence should be addressed to H. Hosono (email: hosono@msl.titech.ac.jp)

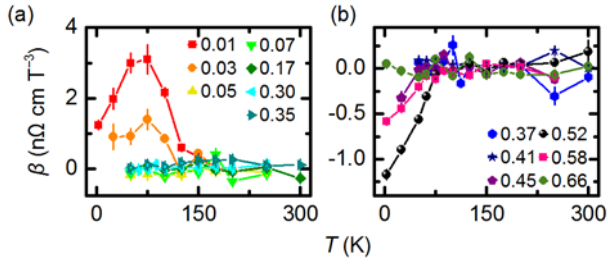

**SM 1.** (a, b) Temperature dependence of the coefficient β in the fit of $\rho_{xy}$ $(H) = R_\text{H}H +$ $\beta H^3 + \gamma$. The data points for $x=0.01$ are multiplied by 1/3. Although at $x = 0.01$ the non-linearity is unobservable above the $T_\text{anom.2} = 138$ K, below the $T_\text{anom.2}$ the β is sharply increased, indicating that this non-linearity comes from the magnetic scattering or reconstruction of the band structure from para-magnet to SDW-type anti-ferro-magnet. Therefore we concluded that except for the temperature region below the $T_\text{anom.2} = 138$, 105, 93, and 100 K at $x = 0.01$, 0.03, 0.52, and 0.58, respectively, the β can be neglected.

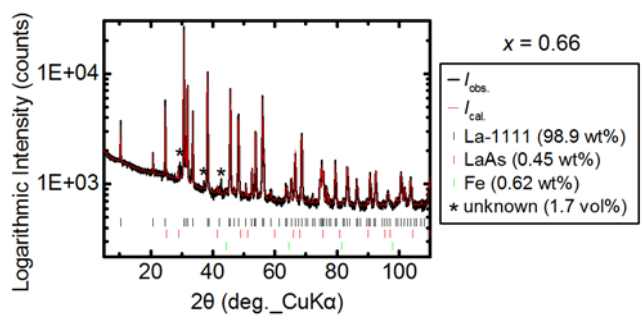

**SM 2.** X-ray diffraction pattern of LaFeAsO$_{0.34}$H$_{0.66}$. The black and red solid lines are observed and calculated intensities. Three bars below the intensities are diffraction positions of LaFeAsO$_{0.34}$H$_{0.66}$ (black), LaAs (red), and Fe (light green) in order from the top. A little amount of unknown phase is detected by XRD and EPMA which is indicated by asterisks. The volume fraction of this phase is estimated as 1.7 %.

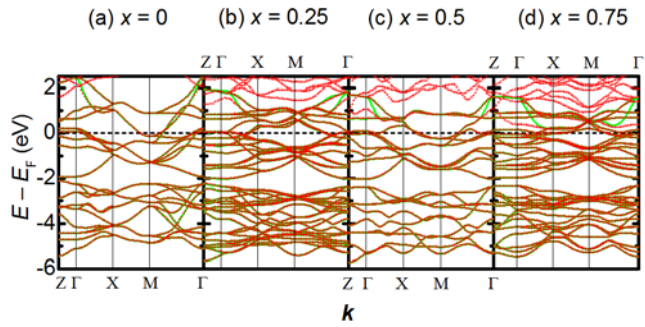

**SM 3.** (a-d) The red symbols and green solid lines indicate the DFT band structures and tight-binding bands obtained from MLWFs including Fe-3*d*, As-4*p*, O-2*p*, and H-1*s* orbitals, respectively. The narrow bands located above and below $E = 2$ eV at $x = 0$ are $5d_{x2-y2}$-orbital of La. The BZs at $x = 0.25$ and 0.75 are folded from the primitive tetragonal BZ of LaFeAsO$_{1-x}$H$_x$ at ambient temperature because we used the $\sqrt{2} \times \sqrt{2} \times 1$ supercell for these calculations.

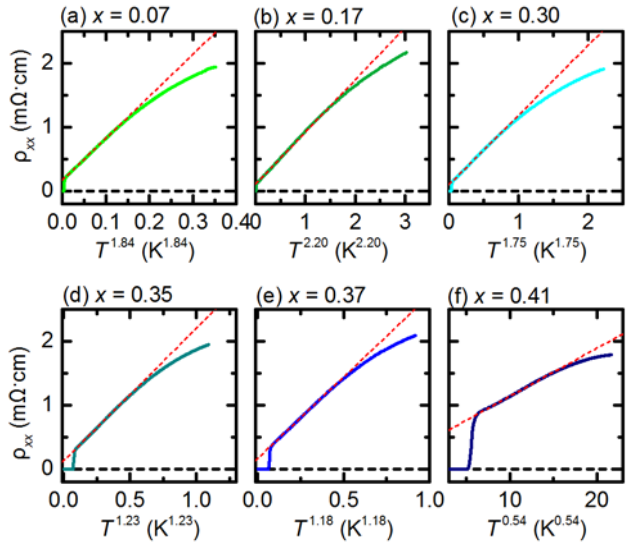

**SM 4.** (a-f) Temperature dependence of resistivity of LaFeAsO$_{1-x}$H$_x$ as a function of $T^n$. The red dased lines are fitting curve, $\rho_{xx}(T) = \rho_{xx}(0) + AT^n$. The temperature range for the fitting is $50 \le T \le 150$ K.

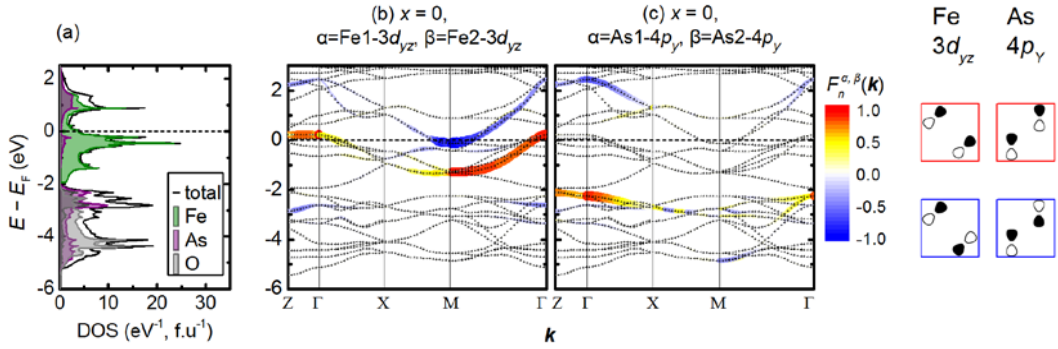

**SM 5.** (a) The total density of states (black solid line) and partial DOS of Fe (green), As (purple), and O (grey) calculated using Slater-Koster-interpolation based on the Wannier orbital basis. (b, c) The band structure colored by $F_n^{\alpha, \beta}(\boldsymbol{k})$ with $\alpha = 3d_{yz}$ at Fe1 site and $\beta = 3d_{yz}$ at Fe2 site (b), and by $F_n^{\alpha, \beta}(\boldsymbol{k})$ with $\alpha = 4p_y$ at As1 site and $\beta = 4p_y$ at As2 site (c). The dashed line denotes the Fermi level. On the right side of the SI 2(c) orbital configurations inside the unit cell are depicted; the red cells represent that the $F_n^{\alpha, \beta}(\boldsymbol{k})$ is positive while the blue that the $F_n^{\alpha, \beta}(\boldsymbol{k})$ is negative. Note that the $3d_{yz}$ orbital shown in the right side of the SI 2(c) is top view, and we choose the $x$ and $y$ axes of the As-orbital along with the crystallographic tetragonal axes, which are different from those of Fe's orbital.